\documentclass[twocolumn,showpacs,preprintnumbers,amsmath,amssymb]{revtex4}

\usepackage{graphicx}
\usepackage{dcolumn}
\usepackage{bm}

\begin{document}

\title{A phase space study of jet formation in planetary-scale fluids}

\author{R.~D. Wordsworth}
 \email{rwlmd@lmd.jussieu.fr}
\affiliation{Laboratoire de M\'et\'erologie Dynamique, Institut Pierre Simon Laplace, Paris, France}

\date{\today}

\begin{abstract}
The interaction between planetary waves and an arbitrary zonal flow is studied from a phase-space viewpoint. Using the Wigner distribution, a planetary wave Vlasov equation is derived that includes the contribution of the mean flow to the zonal potential vorticity gradient. This equation is applied to the problem of planetary wave modulational instability, where it is used to predict a fastest growing mode of finite wavenumber. A wave-mean flow numerical model is used to test the analytical predictions, and an intuitive explanation of modulational instability and jet asymmetry is given via the motion of planetary wavepackets in phase space. 
\end{abstract}
\maketitle

\section{Introduction}\label{sec:intro}
The formation and maintenance of large-scale jets by the collective nonlinear interaction of eddies is a fundamental problem in planetary fluid dynamics. Perhaps the most famous example of the effect can be seen in the atmospheres of the planets Jupiter and Saturn; it has long been believed that the stable, coherent jets observed there owe their existence to continual forcing by smaller-scale eddies.
 Other geophysical examples of interest include the Earth's atmospheric jet stream and, quite possibly, the alternating zonal jets recently observed in the Earth's Pacific ocean \cite{obsoceanjets}. Further afield, an analogous effect involving plasma drift waves is also known to be of great importance in fusion tokomaks \cite{diamond}.
 
In all the geophysical cases, the change of planetary vorticity with latitude or $\beta$-effect is believed to be a vital part of the problem, as it allows for the presence of planetary waves in the system. In a seminal paper, Rhines \cite{rhines} studied the interaction between planetary waves and turbulence, and came to the conclusion that the inverse energy cascade of idealised two-dimensional turbulence would be halted by planetary wave motion at large scales, leading to the transfer of energy into the zonal modes and hence to jet formation. Although extremely insightful, Rhines' work was partly heuristic, and could not provide a detailed dynamical explanation of the process.

As a result, the theoretical investigation of reduced problems involving the interaction of zonal jets and planetary waves is still of great importance to our overall understanding of atmospheric and oceanic fluid dynamics. Much interesting work has previously been done on the subject; for example, a number of authors have studied the interactions of planetary waves with \emph{critical layers} in various idealised scenarios \cite{stewartson}\cite{warnwarn}\cite{killworth}. The foundation for many of these studies was the earlier development of various real-space conservation theorems (see e.g., \cite{charney_drazin}, \cite{andrews_1976}), most of which are now regarded as an essential part of wave-mean flow theory.

Other studies have made use of phase-space transport (Vlasov or Boltzmann) equations to describe the wave-mean flow interaction. Among the first researchers to take this approach were Dyachenko et~al. \cite{dyachenko}, who derived an equation for the interaction between waves and large-scale vortices. Later, Manin \& Nazarenko \cite{manin} used a Vlasov equation to study the interaction between scale-separated zonal flows and planetary waves in the limit of small $\beta$-effect. They found that planetary waves in the presence of a zonal flow could become modulationally unstable, which led to singularity formation and soliton propagation in the model they used.

Recently, there have also been some attempts to utilise phase-space techniques in wave -- mean flow numerical simulations. In Laval et~al. \cite{laval}, scale-separated 2D turbulence was simulated using a Particle-In-Cell (PIC) approach. They treated the small-scale field as an ensemble of `quasiparticles' with a simple dispersion relation $\omega = \overline{\mathbf u} \cdot \mathbf k$ determined solely by the large-scale velocity field $\overline{\mathbf u}$. They compared their method with a direct numerical simulation, and found that it was significantly more computationally efficient.

In comparison to other approaches, however, phase-space techniques have not so far been widely used to study $\beta$-plane jet formation. One reason for this may be that often, the derived transport equations are not compatible with existing real-space results. In addition, the intuitive aspects of the phase space view have not always been clearly emphasized, and no previous studies appear to have tested the predictions of planetary Vlasov equations against more general numerical simulations. In spite of this, the Vlasov formulation offers distinct advantages, as it allows one to build a more  complete picture of interactions between arbitrary distributions of planetary waves and the mean flow than is possible with other methods. It is particularly suited to problems involving collective wave-mean flow instability, such as that considered in Section \ref{sec:compsim2} of this paper.

Here, a new Vlasov equation is derived that describes the interaction of an arbitrary mean flow with a broadband distribution of scale-separated planetary waves. An operator-based derivation is used that allows all effects of the mean flow on the planetary waves to be included for the first time. It is shown that previous real-space results in wave-mean flow theory can be generalised by integrating over the Vlasov equation in spectral space. A numerical simulation is then introduced and used to study some simple but insightful test cases involving the quasilinear motion of a planetary wavepacket. The Wigner distribution of the planetary wavefield is calculated, compared with scale-separated predictions, and used to interpret the simulation results.

Next, the modulational instability analysis of Manin \& Nazarenko is generalised to include the additional potential vorticity effects of the mean flow on the waves. It is found that this generalisation qualitatively changes the dispersion relation for the unstable modes. The numerical simulation is used to test the modified dispersion relation, and the development of the system beyond the initial linear growth phase is also briefly considered. In addition, it is shown that the instability process can be interpreted as a direct result of the motion of wavepackets in phase space.

In Section \ref{sec:theory1}, important results in (real-space) wave-mean flow theory for quasigeostrophic flows are reviewed. In Section \ref{sec:theory2}, the Wigner distribution is defined and used to derive a Vlasov equation for the waves. In Section \ref{sec:compsim}, the numerical simulation is introduced. Finally, in Section \ref{sec:compsim2}, the modulational instability of a planetary wave is studied using both the numerical model and the theoretical methods developed earlier.

\section{Fundamentals of wave-mean flow theory}\label{sec:theory1}
The key features of many large-scale geophysical flows can be captured by the quasigeostrophic potential vorticity (QGPV) equation
\begin{equation}\label{eq:QGPV}
\frac{D q}{D t} = \frac{\partial q}{\partial t}  + J[\psi,q]=-\kappa q
\end{equation}
where $\psi$ is the velocity streamfunction, $q = \left( \partial_{xx} + \partial_{yy} +  \partial_z \left( \left( f_0^2 \slash N^2\right) \partial_z \right)\right) \psi + \beta y $ is the quasigeostrophic potential vorticity, $\beta$ is the local gradient of planetary vorticity with latitude, $J[,]$ is the two-dimensional Jacobian operator such that $J[f,g]=\partial_x f \partial_y g - \partial_x g \partial_y f$ in Cartesian co-ordinates, and $\kappa$ is the Ekman damping parameter. In the definition of $q$, the constant $f_0$ is the Coriolis parameter and $N$ is the buoyancy frequency, which in standard quasigeostrophic theory is a function of $z$ only.

\begin{figure}
 	\begin{center}
 		{\includegraphics[width=3.5in]{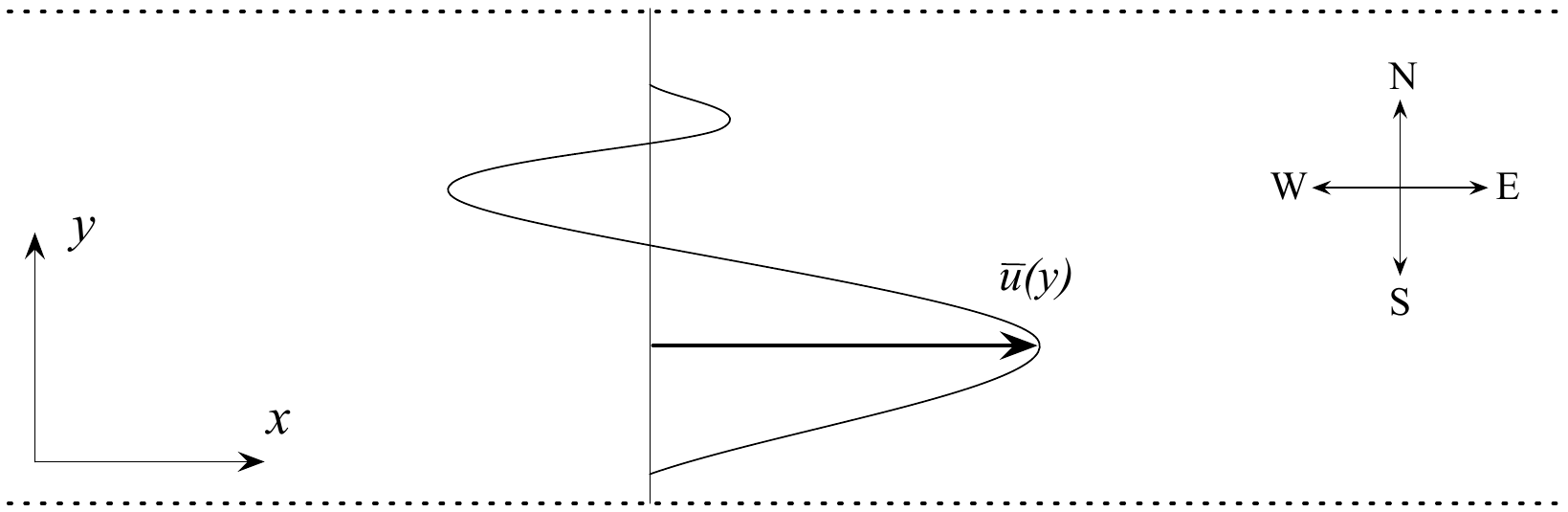}}
 	\end{center}
 	\caption{Schematic of the theoretical setup: a $\beta$-plane model periodic in the $x$-direction and open in the $y$-direction. The $\beta$-plane approximates fluid motion on the midlatitudes of a planet, with $x$ and $y$ equivalent to east-west and north-south directions respectively. }\label{fig:betachan}
\end{figure}

Equation (\ref{eq:QGPV}) is simply a statement that $q$ is conserved following fluid elements in the absence of damping and forcing. It is an approximation to the full Navier--Stokes equations that is applicable when the system under consideration is rapidly rotating and strongly stratified, and is derived in detail in many fluid dynamics textbooks (see e.g., \cite{salmon}). For convenience, in this paper we work with (\ref{eq:QGPV}) in Cartesian co-ordinates according to the standard $\beta$-plane model (see Figure \ref{fig:betachan}). We also mostly focus on the situation where the system is unbounded in the $y$ (north-south) direction, although in Section \ref{sec:compsim2}, the north-south boundary conditions are set to be periodic for simplicity. The theoretical setup is summarised in Figure \ref{fig:betachan}.

As a result of the background vorticity gradient $\beta$, the linearised form of (\ref{eq:QGPV}) has planetary wave solutions, with dispersion relation
\begin{equation}\label{eq:QGPVwave}
\sigma = \frac{-\beta k_x}{k_x^2+k_y^2+k_z^2}
\end{equation}
when no zonal flow or damping is present. In (\ref{eq:QGPVwave}), $k_z$ is understood to be an eigenvalue of the usual vertical structure equation such that $ \partial_z \left( \left(f_0^2 \slash N^2 \right) \partial_z \Psi \right) = -k_z^2\Psi$, subject to suitable boundary conditions.

To investigate the interaction between planetary waves and zonal flow, it is standard to define an average in the $x$-direction such that any quantity decomposes into a mean flow and a disturbance field: $f(x,y,t) = \overline {f(y,t)}  +  f'(x,y,t)$. Then, (\ref{eq:QGPV}) becomes
\begin{equation}\label{eq:QGPVmean}
\frac{\partial \overline q}{\partial t} = - \frac{\partial}{\partial y} \overline {v'q'}  -\kappa \overline q 
\end{equation}
for the mean flow and
\begin{equation}\label{eq:QGPVdist}
\frac{\partial q'}{\partial t} + \overline u\frac{\partial q'}{\partial x} + \gamma v' = \overline{J[\psi',q']} - J[\psi',q'] - \kappa q'
\end{equation}
for the disturbances, with $\gamma = \partial_y \overline q$ the total gradient of zonal potential vorticity. The left hand side of (\ref{eq:QGPVdist}) describes the evolution of planetary waves in the presence of a zonal flow, while the right describes nonconservative effects and wave-wave interactions.

As this work is primarily concerned with wave-mean flow interaction, we will assume the wave-wave terms in (\ref{eq:QGPVdist}) to be small from here onwards. In Section \ref{sec:compsim}, where we discuss the numerical simulation of a generic jet-wave interaction problem, the situations where this assumption begins to fail will be made clear.

For moderate disturbance amplitudes, it can be shown that the \emph{wave action}, defined as $n\equiv \frac 12 \overline{q'^2}\slash \gamma$, is a conserved quantity. This can be seen through multiplication of  (\ref{eq:QGPVdist}) by $q' \slash \gamma$ and zonal averaging, which results in
\begin{equation}\label{eq:waveaction}
\frac{\partial n}{\partial t} + \overline{v'q'} = -2\kappa n
\end{equation}
if terms of order $q'^3$ and greater are neglected. If spatial scale separation between zonal flow and waves is then assumed and the waves are taken to be monochromatic, a real-space transport equation for wave action can be written
\begin{equation}\label{eq:waveaction_scalesep}
\frac{\partial n}{\partial t} + \nabla_{m} \cdot  \left( \mathbf v_{m} n \right)  = -2\kappa n,
\end{equation}
where $\nabla_{m}=(\partial_y,\partial_z)$  and $\mathbf v_{m}$ is the meridional $(y,z)$ group velocity of the waves. For further details of the derivation of (\ref{eq:waveaction}), see \cite{MAD}.

Note that the definition of $n$ given here depends on the potential vorticity gradient, $\gamma$, remaining non-zero. If $\gamma$ changes sign somewhere in the domain, then the zonal flow may be unstable. The problems associated with defining $n$ in these cases are discussed in more detail in \cite{pedlosky}. 

The final established result of importance to the rest of this paper is the powerful \emph{non-acceleration theorem}, which states that in the absence of forcing or damping, the rate of change of zonal potential vorticity and wave action are directly tied to each other
\begin{equation}\label{eq:nonaccel}
\frac{\partial  }{\partial t}\left(  \overline q  - \frac{\partial n}{ \partial y} \right)  = 0.
\end{equation}
Equation (\ref{eq:nonaccel}) can be proved by use of the Taylor identity \cite[]{MAD}, or by combining equations (\ref{eq:QGPVmean}) and (\ref{eq:waveaction}) and setting $\kappa = 0$.

\section{Derivation of the planetary wave Vlasov equation}\label{sec:theory2}

In this section, we generalise the results reviewed in the previous section to arbitrary broadband distributions of planetary waves. As was mentioned in Section \ref{sec:intro}, equations for broadband wave-mean flow interaction have been used in several previous studies. The aim here is to derive a Vlasov equation that is directly compatible with the real-space wave-mean results reviewed in the previous section. As will be shown, this requires the inclusion of additional mean flow effects that qualitatively change the predictions, particularly in the modulational instability analysis of Section \ref{sec:compsim2}.

 We begin the derivation by writing the disturbance equation (\ref{eq:QGPVdist}) in terms of a new variable $\phi\equiv q' \slash \sqrt{2\gamma}$. If we neglect terms of order $\phi^2$ and higher, (\ref{eq:QGPVdist}) then takes the form
\begin{equation}\label{eq:newvar}
i\frac{\partial \phi}{\partial t} = \hat G \phi ,
\end{equation}
where the wave operator $\hat G$ is defined as
\begin{equation}\label{eq:Gop}
\hat G[\mathbf{\hat x},\mathbf{\hat k},t] \equiv \sqrt{\gamma}\frac{-\hat k_x}{\hat k_x^2+ \hat k_y^2+ \hat k_z^2}\sqrt{\gamma} + \overline u \hat k_x - i\kappa.
\end{equation}
Note that $\hat G$ is a self-adjoint operator when the damping term $\kappa = 0$. This is part of the motivation for introducing the `square root of wave action' variable $\phi$. As will be seen later, our choice of $\phi$ is also very important in ensuring that the new results correctly generalise the real-space wave action equation (\ref{eq:waveaction_scalesep}).

The position and wavevector operators are 
\begin{equation}\label{eq:Wops}
\hat {\mathbf x} = \mathbf x  \qquad \mbox{and}  \qquad \hat {\mathbf k} = \left( -i \partial_x, -i \partial_y, -i\sqrt{\partial_z \left( \left(f_0^2 \slash N^2 \right) \partial_z \right)}\right)
\end{equation}
respectively, as we are working in a position space representation. The denominator in (\ref{eq:Gop}) is simply the potential vorticity inversion operator, such that 
\begin{eqnarray}
\psi &=&-\left( {\hat k_x^2+ \hat k_y^2+ \hat k_z^2}\right)^{-1}q  \nonumber\\
	&=& \left( \partial_{xx}+\partial_{yy} + \partial_z (f_0^2 \slash N^2) \partial_z \right)^{-1}q.
\end{eqnarray}

When scale separation \emph{is} assumed, operators become real numbers, and the large-scale zonal flow `sees' wavepackets as phase-space points with exact values of  $\mathbf x$ and $\mathbf k$. Then, (\ref{eq:Gop}) simply becomes the generalised dispersion relation for small-scale planetary waves in the presence of a zonal flow and damping
\begin{equation}\label{eq:Gdisp}
\hat G[\mathbf{\hat x},\mathbf{\hat k},t] \to \omega(y,z,\mathbf{k},t) = \frac{-\gamma k_x}{k_x^2+ k_y^2+ k_z^2} + \overline u  k_x - i\kappa.
\end{equation}

At this point, we need to utilize a tool from quantum mechanics: the Wigner distribution. It is defined in three dimensions as
\begin{equation}\label{eq:Wigdef}
\mathcal N_{\phi, \phi}(\mathbf x, \mathbf k, t) = \frac1{(2\pi)^3} \int^{+\infty}_{-\infty} \phi^*_{\mathbf x - \frac 12 \mathbf x_1} e^{-i\mathbf k \cdot \mathbf x_1}\phi_{\mathbf x + \frac 12 \mathbf x_1} \mbox{d}^3 \mathbf x_1
\end{equation}
or alternatively in spectral space as
\begin{equation}\label{eq:Wigdef2}
\mathcal N_{\phi, \phi}(\mathbf x, \mathbf k, t) =  \frac1{(2\pi)^3} \int^{+\infty}_{-\infty} \Phi^*_{\mathbf k - \frac 12 \mathbf k_1} e^{i\mathbf x \cdot \mathbf k_1}\Phi_{\mathbf k + \frac 12 \mathbf k_1} \mbox{d}^3 \mathbf k_1 
\end{equation}
where $\Phi(\mathbf k,t) ={(2\pi)}^{-3 \slash 2}\int^{+\infty}_{-\infty} \mbox{exp}[-i\mathbf k \cdot \mathbf x]\phi(\mathbf x,t)\mbox{d}^3 \mathbf x$ is the Fourier transform of $\phi$. $\mathcal N_{\phi,\phi}$ can broadly be thought of as a phase-space distribution for the function $\phi$, but it has some fairly weird properties --- not least of which being that it can take \emph{negative} values. However, its projections onto real and spectral space are always positive valued. 

By taking a time derivative of (\ref{eq:Wigdef}) and using (\ref{eq:newvar}), we may write
\begin{equation}\label{eq:Wigtime}
i\frac{\partial \mathcal N_{\phi, \phi}}{\partial t} =  \mathcal N_{-\hat G \phi, \phi} +  \mathcal N_{\phi, \hat G \phi}.
\end{equation}
Then, by defining the \emph{phase-space} operators $\mathbf{\hat X} = \mathbf x +  \frac i2 \nabla_{\mathbf k}$ and $\mathbf{\hat K} =\mathbf k - \frac i2 \nabla_{\mathbf x}$ (see e.g., \cite{torre})
and noting that 
\begin{eqnarray}\label{eq:WigkOp}
\mathbf{\hat X} \mathcal N_{\phi,\phi} =  \mathcal N_{\phi,\mathbf{\hat x}\phi} \qquad
\mathbf{\hat K} \mathcal N_{\phi,\phi} =  \mathcal N_{\phi,\mathbf{\hat k} \phi},
\end{eqnarray}
and hence clearly
\begin{eqnarray}\label{eq:WigkOp2}
\mathbf{\hat X}^n \mathcal N_{\phi,\phi} =  \mathcal N_{\phi,\mathbf{\hat x}^n\phi} \qquad
\mathbf{\hat K}^n \mathcal N_{\phi,\phi} =  \mathcal N_{\phi,\mathbf{\hat k}^n \phi},
\end{eqnarray}
the fairly weak assumption that $\hat G[\mathbf{\hat X},\mathbf{\hat K},t]$ can be expanded in powers of the two operators $\mathbf{\hat X}$ and $\mathbf{\hat K}$ (i.e., that it has a valid Taylor series representation) allows us to arrive at
\begin{equation}\label{eq:Wigtrans}
i \frac{\partial \mathcal N_{\phi,\phi} }{\partial t} = \left( \hat G[\mathbf{\hat X},\mathbf{\hat K},t] - \hat  G[\mathbf{\hat X^*},\mathbf{\hat K^*},t] \right) \mathcal N_{\phi,\phi}.
\end{equation}
Equation (\ref{eq:Wigtrans}) is the Wigner transport equation, describing the motion of planetary wave action in phase space without an assumption of scale separation. Although it is general, its form is based on operator notation, which unfortunately makes direct analysis difficult. Therefore, we derive the Vlasov equation via a Taylor expansion of the operator $\hat G$ about $\mathbf x$ and $\mathbf k$. Truncation of the expansion at first order allows us to write
\begin{equation}\label{eq:WigExpand}
\hat G[\mathbf{\hat X},\mathbf{\hat K},t] \approx \omega(\mathbf x,\mathbf k,t)+ \frac{\partial \omega}{\partial \mathbf x} \cdot \frac i2 \frac{\partial}{\partial \mathbf k}- \frac{\partial \omega}{\partial  \mathbf k}\cdot \frac i2 \frac{\partial}{\partial \mathbf x}.
\end{equation}
Provided that such a representation for $\hat G$ is valid, a similar expansion for $\hat G[\mathbf{\hat X^*},\mathbf{\hat K^*},t]$ and substitution into (\ref{eq:Wigtrans}) then leads to
\begin{equation}\label{eq:WKstd}
\frac{\partial \mathcal N}{\partial t} + \mathbf v \cdot \frac{\partial \mathcal N}{\partial \mathbf x} +
 \mathbf F \cdot \frac{\partial \mathcal N}{\partial \mathbf k} = \Gamma[\mathcal N]
\end{equation}
where $\mathbf v$ and $\mathbf F$, the group velocity of, and force on, a wavepacket respectively, have their usual definitions as $\mathbf v = \nabla_{\mathbf k} \omega$ and $\mathbf F = -\nabla_{\mathbf x} \omega$. For brevity, we write $\mathcal N_{\phi,\phi} = \mathcal N$ from here.

Equation (\ref{eq:WKstd}) is equivalent to (\ref{eq:Wigtrans}) in the geometrical optics limit of small-scale disturbances. It describes the collective motion of an ensemble of point-like wavepackets through phase space. The right hand side of (\ref{eq:WKstd}) contains all nonconservative terms: according to our derivation, $\Gamma[\mathcal N] = -2\kappa \mathcal N$. However, if the effects of wave-wave interactions were to be included, $\Gamma$ would also contain more complicated terms describing collisions between wavepackets. For the general case of planetary waves on an arbitrary zonal flow, these terms are not known. They have been derived for small-scale planetary wave interaction in the \emph{absence} of zonal flow by several authors, beginning with Longuet-Higgins \& Gill \cite{longuet}. As the interaction of planetary waves is well-known to be incapable of giving energy to the zonal flow at lowest order, and our main interest is the interaction between waves and zonal flow, we will not make use of these results here.

Differentiation of (\ref{eq:Gdisp}) in phase space yields the group velocities and rates of change of wavenumber or `forces' on planetary wavepackets
\begin{eqnarray}\label{eq:WKgroup}
 \mathbf v &=&  \left( -\frac{\gamma}{|\mathbf k|^2} +\overline u + \frac{2\gamma k_x^2}{|\mathbf k|^4}, \frac{2 \gamma k_xk_y}{|\mathbf k|^4} , \frac{2 \gamma k_xk_z}{|\mathbf k|^4}  \right) \nonumber \\
 \mathbf F &=& \left( 0, \frac{  k_x}{|\mathbf k|^2}\frac{\partial  \gamma}{\partial y} - k_x \frac{\partial \overline u}{\partial y} , 
 \frac{  k_x}{|\mathbf k|^2}\frac{\partial  \gamma}{\partial z} - k_x \frac{\partial \overline u}{\partial z}
\right),
\end{eqnarray}
with $|\mathbf k|^2=k_x^2+k_y^2+k_z^2$ and  $\mathbf v = \mbox{d} \mathbf x \slash \mbox{d} t$, $\mathbf F = \mbox{d} \mathbf k \slash \mbox{d} t$.

Finally, if we \emph{define} the wave action $n$ (see equation (\ref{eq:waveaction})) to be the projection of $\mathcal N$ onto $(y,z)$ real space
\begin{equation}\label{eq:Nandn}
n(y,z,t)  \equiv \overline{\int^{+\infty}_{-\infty} \mathcal N(\mathbf x,\mathbf k, t) \mbox{d}^3 \mathbf k},
\end{equation}
where the overline denotes the zonal average defined earlier, we can extend the results of standard wave-mean flow theory outlined in Section \ref{sec:theory1}. By integrating (\ref{eq:WKstd}) over $\mathbf k$, making use of the fact that $\nabla_{\mathbf x} \cdot \mathbf v +\nabla_{\mathbf k} \cdot \mathbf F=0$ and assuming that $\mathcal N \to 0$ as $|\mathbf k|\to \infty$, 
we arrive at 
\begin{equation}\label{eq:waveaction_scalesep_gen}
\frac{\partial n}{\partial t} + \nabla_{2D} \cdot \left( \langle \mathbf v_{2D} \rangle n \right) = -2\kappa n, \end{equation}
where 
\begin{equation}\label{eq:ensemble_vel}
\langle \mathbf v_{2D} \rangle \equiv \frac 1n \overline{\int^{+\infty}_{-\infty} \mathbf v \mathcal N \mbox{d}^3 \mathbf k}
\end{equation}
is the spectrally averaged $(y,z)$ group velocity for the planetary waves. Equation (\ref{eq:waveaction_scalesep_gen}) is a generalisation of (\ref{eq:waveaction_scalesep}) to a broadband distribution of small-scale waves, which is made possible by the initial definition of $\phi$, not $q'$, as the quantity in the wave equation (\ref{eq:newvar}). Interestingly, its derivation from (\ref{eq:WKstd}) is closely analogous to the derivation of the continuity equation from the Boltzmann equation in fundamental fluid dynamics. 

\section{Numerical simulation I: Motion of a single wavepacket}\label{sec:compsim}
We now wish to develop a more intuitive understanding of the ideas of the previous section, by considering a wave-mean flow numerical simulation. The simple test cases studied here are interesting in their own right, but they are also important for the phase-space interpretation of planetary wave modulational instability, which is discussed in Section \ref{sec:compsim2}.

The essential features of the wave-mean flow problem are captured by restricting the planetary wavefield to a single east-west wavenumber, $k_x=k_0$, but allowing it to be broadband in $k_y$. 
This is justified by noting that according to (\ref{eq:WKgroup}), the zonal flow cannot move the planetary wavepackets to different $k_x$ and also that in a wave-mean flow context, their $x$-position is clearly irrelevant. For simplicity we also ignore damping ($\kappa=0$), and restrict the problem to the single layer \emph{barotropic} case. However, it should be noted that all of the theory presented in the previous section is also applicable to mixed barotropic / baroclinic flows, which in general will vary with height as well as latitude and longitude. 

For barotropic planetary waves, only the $y$ and $k_y$ dimensions of phase space are of relevance. In particular, the equation for phase-space velocity vectors (\ref{eq:WKgroup}) simplifies to the two components
\begin{eqnarray}\label{eq:WKgroup2D}
v_y = \frac{2 \gamma k_xk_y}{|\mathbf k|^4}, \qquad
F_y = \frac{  k_x}{|\mathbf k|^2}\frac{\partial  \gamma}{\partial y} - k_x \frac{\partial \overline u}{\partial y}.
\end{eqnarray}

To investigate equations (\ref{eq:QGPVmean}) and  (\ref{eq:QGPVdist}) numerically, we write the disturbance vorticity as $q'=\mbox{Re}[Qe^{ik_0 x}]$, allowing the derivation of the simplified equations
\begin{equation}\label{eq:code1}
i\frac{\partial Q}{\partial t} = k_0\left( \overline u Q + \gamma \Psi \right)-i\kappa Q \qquad \hat \Psi \equiv -\left(\hat k_y^2 + k_0^2 \right)^{-1}Q
\end{equation}
and
\begin{equation}\label{eq:code2}
\frac{\partial \overline u}{\partial t} = \overline{v'q'}- \kappa \overline u=-\frac{k_0}2 \left(\mbox{Im}[\Psi]\mbox{Re}[Q] -  \mbox{Im}[Q]\mbox{Re}[\Psi]   \right) - \kappa \overline u
\end{equation}
for waves and mean flow respectively.  It should be emphasized at this point that $\hat k_y=-i\partial_y$ is an operator, as defined in (\ref{eq:Wops}), and hence (\ref{eq:code1}) and (\ref{eq:code2}) make no assumption of scale separation.

For all the numerical results presented here, (\ref{eq:code1}) and (\ref{eq:code2}) were solved using an explicit 4\textsuperscript{th} order Runge-Kutta method. The program was designed to halt whenever a) the Rayleigh-Kuo criterion $\beta-\overline u'' < 0$ for barotropic instability or b) the heuristic wave-breaking criterion $|u'|_{max}>\omega \slash k_x$ were satisfied. This ensured that the original physical assumptions behind the model were not broken during the simulation.

As this simulation is highly idealised and not intended to directly model real planetary flows, dimensionless units are used throughout this section. For comparison, however,  we note that for a midlatitude slice of Jupiter's atmosphere, when scaled into units of planetary rotation period $T_J$ and radius $r_J$, the mean zonal wind speed is approximately $\overline u=0.01\mbox{ }r_J\mbox{ }T_J\mbox{ }^{-1}$ and the $\beta$ parameter is $\beta = 5-10 \mbox{ }r_J\mbox{ }^{-1}\mbox{ }T_J\mbox{ }^{-1}$, depending on latitude. For all simulations in this section we used $\beta = 10$, and maximum zonal wind speeds of barotropically stable jets were of order $\mbox{max}[\overline u] = 0.001$.  Thus we are investigating a fluid dynamical regime with slightly weaker zonation, generally, than that observed on the gas giant planets.

\begin{figure}
 	\begin{center}
 		{\includegraphics[width=3in]{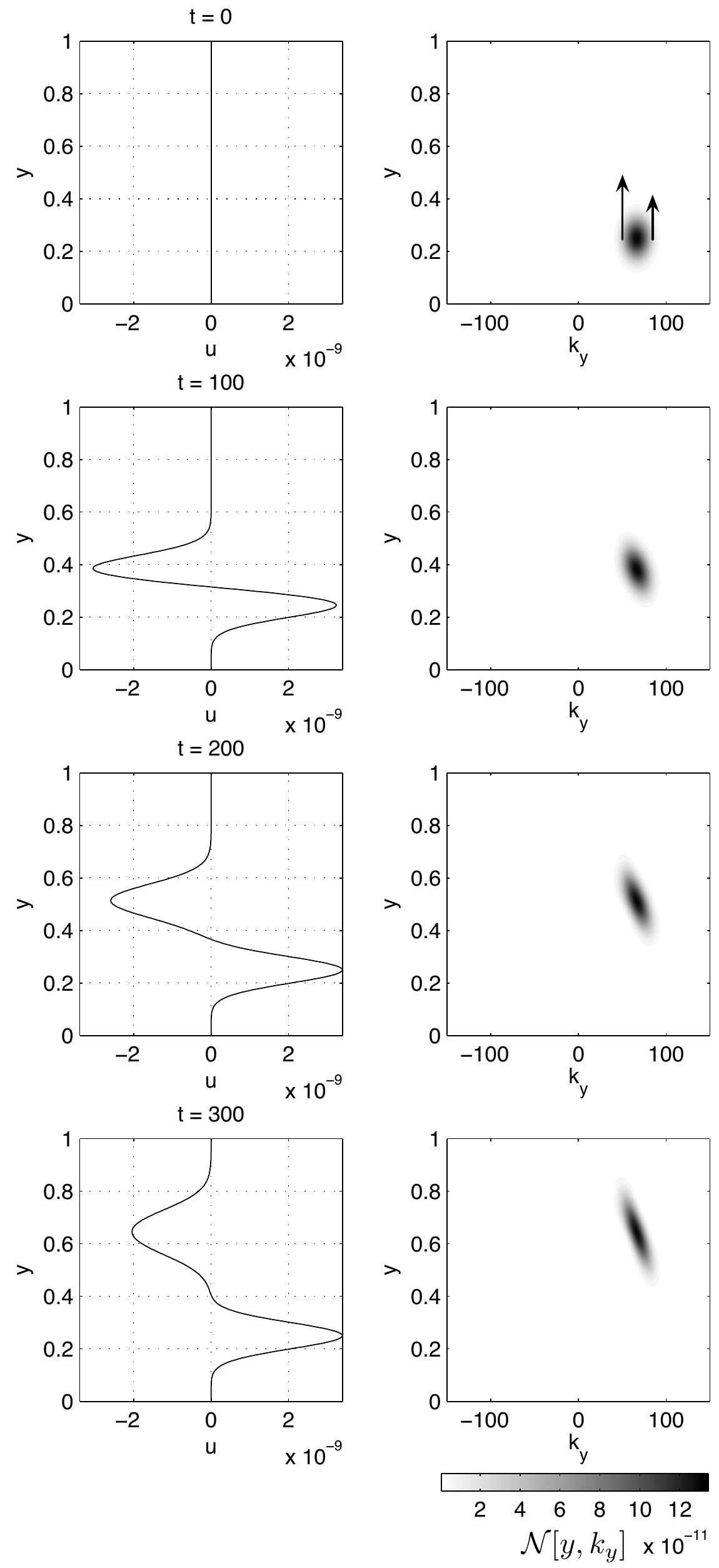}}
 	\end{center}
 	\caption{A planetary wavepacket on a $\beta$-plane with positive $k_y$ and $k_y$ will move northwards. As group velocity depends on $k_y$, the wavepacket becomes tilted in phase space (second column), although its volume remains approximately the same. Note the small-amplitude zonal flow (first column) induced by the wavepacket motion.}\label{fig:beta_wp}
\end{figure}

First, we study an extremely simple test case: a near-infinitesimal wavepacket with no initial zonal flow and no Ekman damping. The initial disturbance vorticity is 
\begin{equation}
Q = Q_0\mbox{exp}\left[{il_0y -(y-y_0)^2 \slash (\Delta y)^2} \right],
\end{equation}
with $l_0=k_0=60$, $y_0=0.25$ and $\Delta y = 0.1$. In Figure \ref{fig:beta_wp}, the magnitude of the Wigner distribution $|\mathcal N_{\phi,\phi}|$ is plotted above zonal velocity $\overline u$ for a series of timesteps. For all the plots in this section, $\mathcal N_{\phi,\phi}$ was calculated directly from the numerical eddy vorticity, without any scale separation assumption.  

\begin{figure}
 	\begin{center}
 		{\includegraphics[width=3.5in]{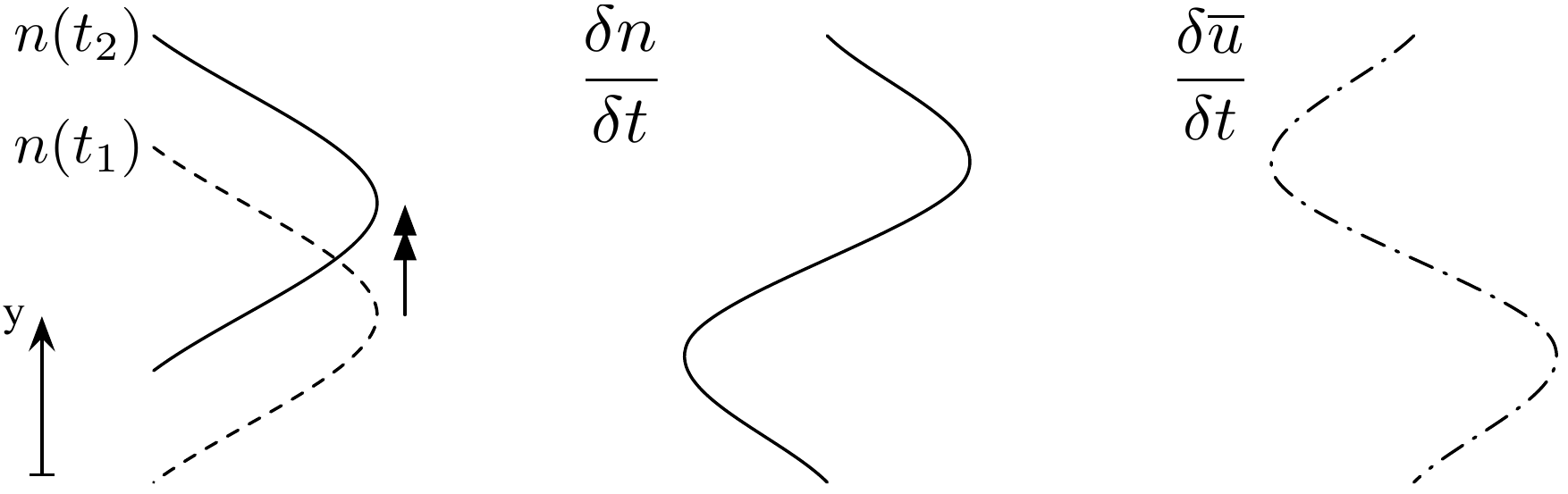}}
 	\end{center}
 	\caption{Schematic explanation of the jet formation seen in Figure \ref{fig:beta_wp}. If a planetary wavepacket is moving northwards such that in time $\delta t=t_2-t_1$, $\delta n = n(t_2) - n(t_1)$, then the nonacceleration theorem (\ref{eq:nonaccelbaro}) ensures the zonal flow produced $\delta \overline u$ will be of the form shown.}\label{fig:nonaccel}
\end{figure}

As can be seen, when the wavepacket has wavevector such that $k_x k_y > 0$, it drifts northwards due to the $\beta$-effect. Weak zonal jets form as a result of this motion. By the barotropic version of the nonacceleration theorem (\ref{eq:nonaccel})
\begin{equation}\label{eq:nonaccelbaro}
\frac{\partial \overline u}{\partial t}=-\frac{\partial n}{\partial t}
\end{equation}
we see that latitudinal planetary wavepacket motion must always cause jets to form in this way. This process is summarised in Figure \ref{fig:nonaccel}. 

The group velocity calculated from (\ref{eq:WKgroup2D}) agrees closely with the observed velocity of the wavepacket peak (the difference is less than 3$\%$ in the example shown). However, note the stretching of the wavepacket in phase space, due to the  dispersive nature of the planetary waves, as determined by (\ref{eq:QGPVwave}). Essentially, the local group velocity $v_y$ on the left hand side (in phase space) of the wavepacket is greater than that on the right --- this is shown by the arrows on the first plot in Figure \ref{fig:beta_wp}.

\begin{figure}
 	\begin{center}
 		{\includegraphics[width=3in]{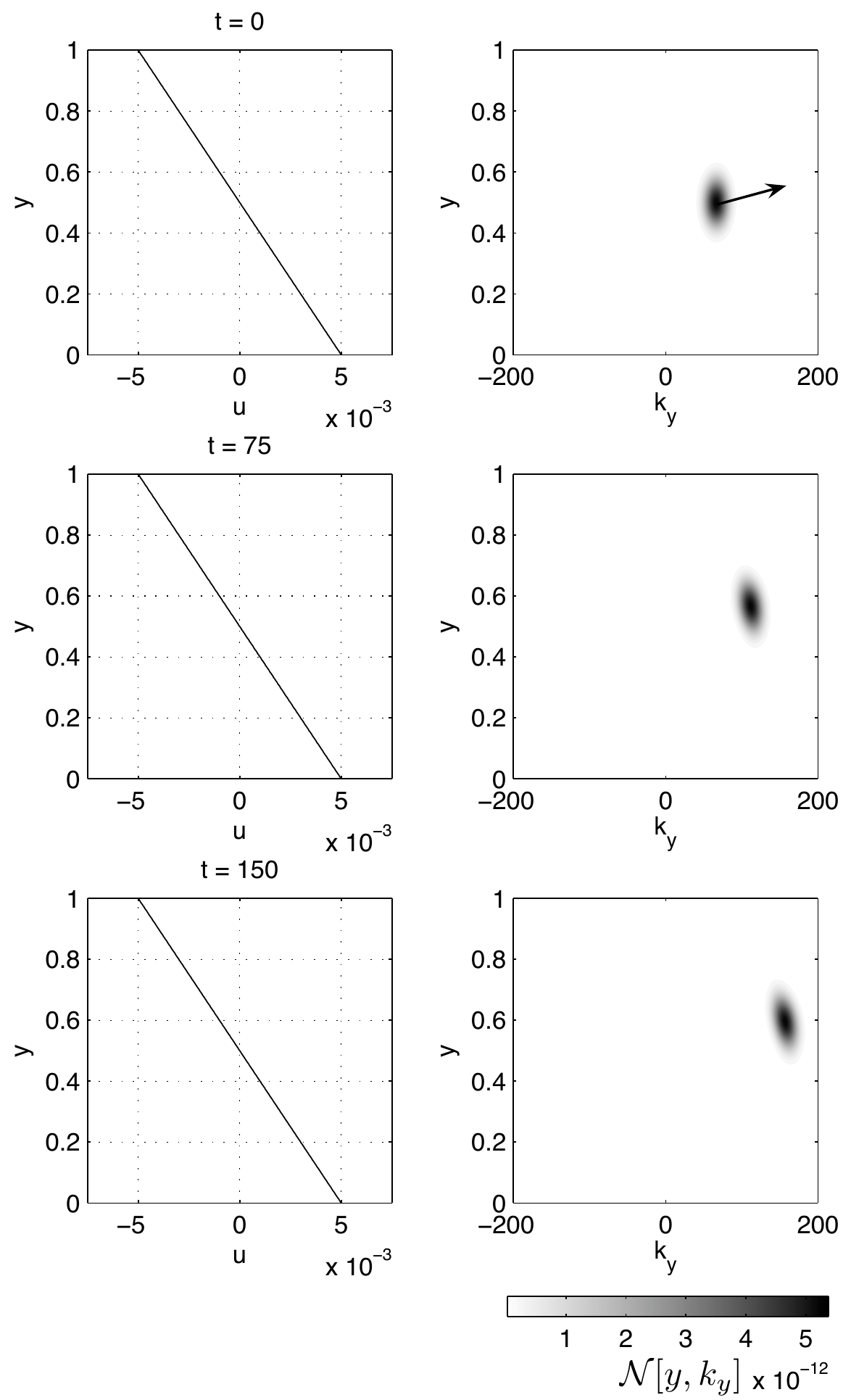}}
 	\end{center}
 	\caption{In the presence of a shear flow that decreases linearly with latitude (far left), a planetary wavepacket with positive $k_x$ will move toward higher wavenumbers, losing energy to the zonal flow in the process. Note the slight northward drift of the wavepacket; this is due to the $\beta$-effect shown in Figure \ref{fig:beta_wp}.}\label{fig:shear}
 \end{figure}

The second basic case of interest involves an infinitesimal wavepacket on a linearly sheared zonal flow of the form $\overline u= - \Lambda (y-y_0)$. Here, $\Lambda = 0.01$, $y_0=0.5$ and all other parameters are as in the previous example. As shown in Figure \ref{fig:shear}, in this situation a wavepacket with $k_x>0$ is forced towards higher $k_y$ wavenumbers, losing energy to the zonal flow in the process. As $\beta=\gamma$ in this example, the enstrophy of the wavepacket remains constant and hence energy is transferred upscale, while enstrophy moves downscale. Again, scale separated predictions match the observed value closely for this case.

\begin{figure}
 	\begin{center}
 		{\includegraphics[width=3in]{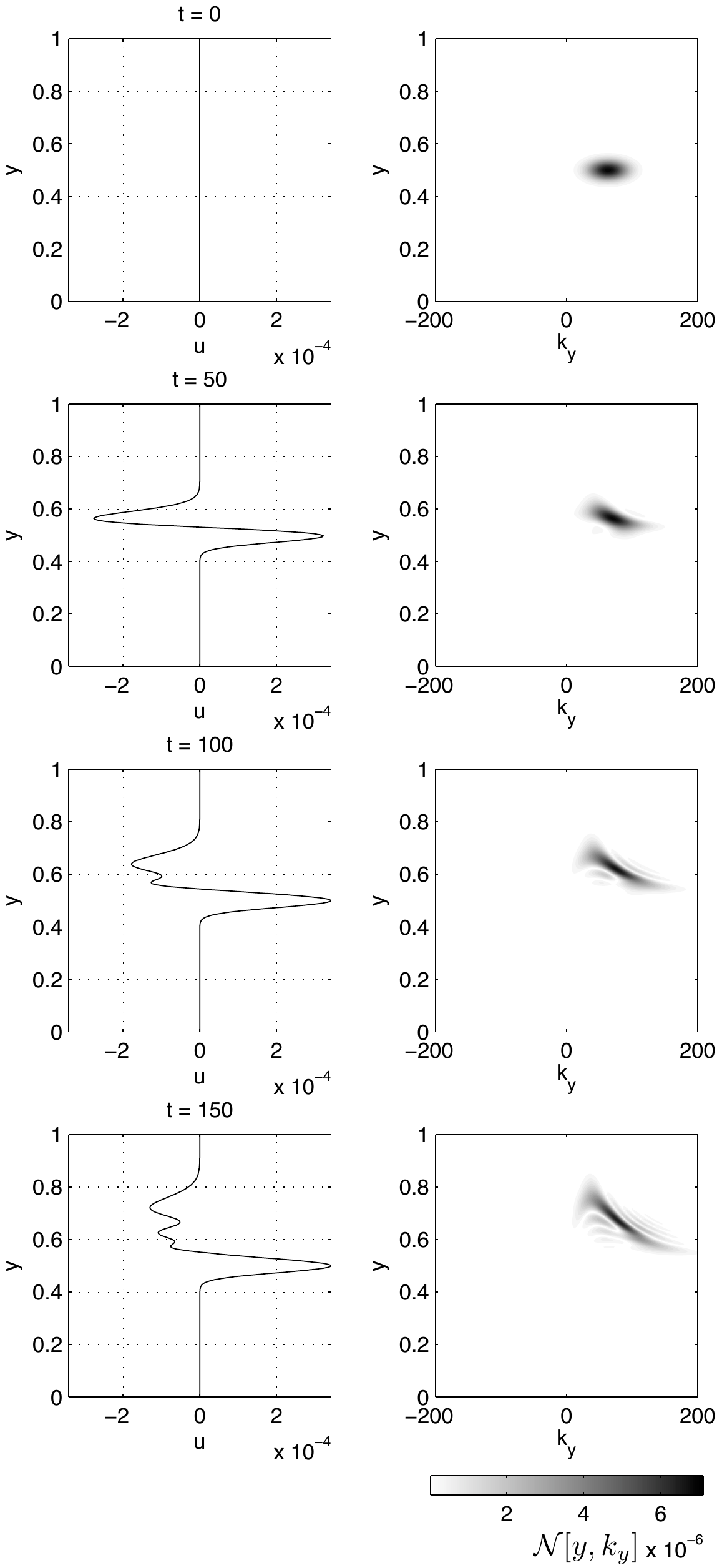}}
 	\end{center}
 	\caption{Quasilinear evolution of a wavepacket ($Q_0=0.12$, $y_0=0.5$, $\Delta y=0.05$, all other parameters as in first example). Initially, the wavepacket moves as in Figure \ref{fig:beta_wp}, but the zonal shear it produces modifies its motion as time progresses. Note that by $t = 100$, the Wigner distribution has become negative-valued in places.}\label{fig:coupled1}
 \end{figure}

More interesting and subtle phenomena occur when we allow the wavepacket to be of large enough amplitude for coupling with the zonal flow, but not wave-breaking, to occur (see Figure \ref{fig:coupled1}). Then, we expect it to initially move northwards, with two zonal jets forming due to the non-acceleration theorem, as in the first example. However, as the zonal flow becomes stronger, it begins to influence wavepacket propagation through a) the shear effect described in the second example and b) alteration of the basic potential vorticity gradient $\gamma = \beta - \partial_{yy} \overline u$ (see equation (\ref{eq:WKgroup2D})). As the initial zonal flow gradient in the centre of the channel is negative, the wavepacket is forced to higher $k_y$ wavenumbers, reducing its group velocity and hence the growth rate of the zonal flow. This process continues until the zonal flow either removes most of the wavepacket energy and reaches a quasi-steady state, or sharpens to the extent that it becomes barotropically unstable.

 \begin{figure}
 	\begin{center}
 		{\includegraphics[width=3in]{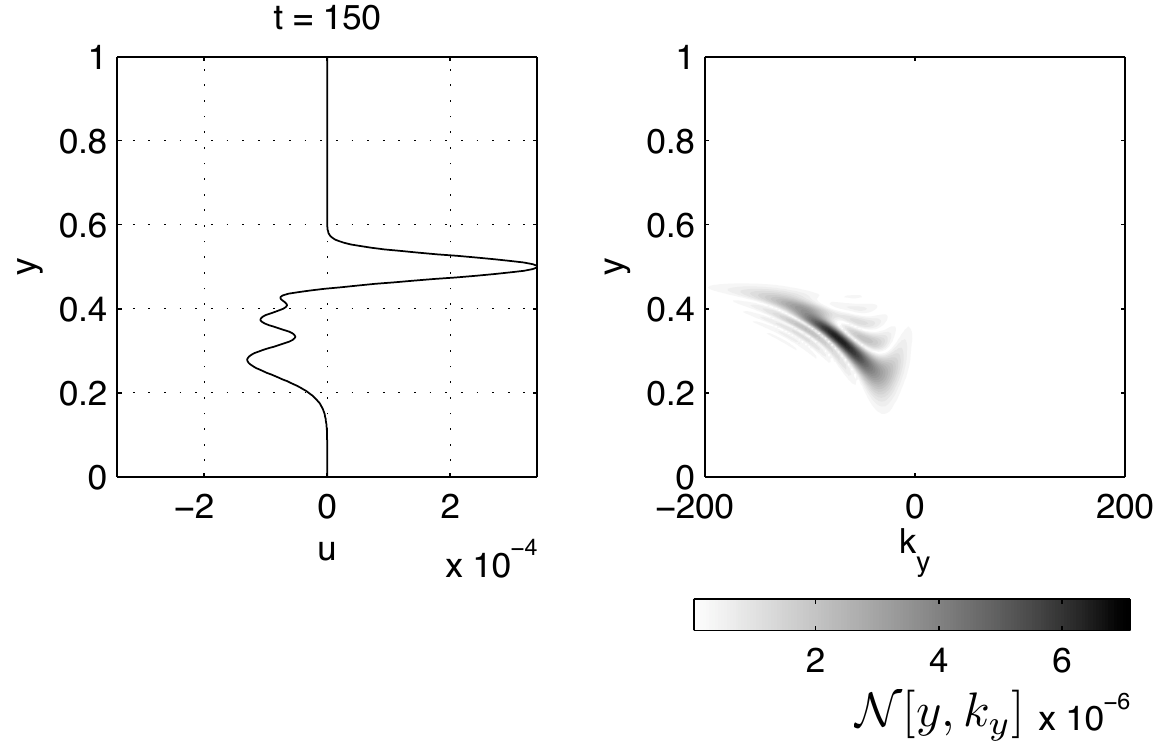}}
 	\end{center}
 	\caption{State at $t=150$ of the same simulation as Figure \ref{fig:coupled1}, but with the initial $k_y$ value of the wavepacket reversed.
	}\label{fig:coupled2}
\end{figure}

Interestingly, the same east-west jet asymmetry occurs regardless of the wavevector $\mathbf k$ sign of the initial wavepacket. As expected from (\ref{eq:WKgroup2D}) and shown in Figure \ref{fig:coupled2}, if the product $k_x k_y$ is negative, the wavepacket initially moves south, and the initial induced jets are of opposite sign. However, whatever the sign of $k_x$ and $k_y$, jet formation always pushes some wave action to higher \emph{absolute} wavenumber values $|k_y|$. Combined with the fact that the wavepacket propogates away from the jet in a direction dependent on $k_xk_y$, the result is that in each case, the eastward jet becomes sharper than the westward one.

 Evidence of quasigeostrophic jet asymmetry has been found in several previous numerical studies (e.g., \cite{haynes},  \cite{chekhlov}, \cite{leesmith}). Indeed, east-west asymmetry appears to be a quite generic feature of wave-forced jets on the $\beta$-plane. The examples given here simply demonstrate the intuitive explanation of the phenomenon that is possible from a phase-space viewpoint.

\section{Numerical simulation II: Modulational instability}\label{sec:compsim2}

In this section, we use (\ref{eq:WKstd}) and the numerical model just described to study the important problem of planetary-wave modulational instability. While it  has been known in principle that systems of planetary waves are unstable to modulations since at least the study of Newell \cite{newell} (see also the related Gill \cite{gill74}), the phenomenon has received less attention to date in planetary fluid dynamics than it perhaps deserves.  In Esler \cite{esler}, the \emph{longitudinal} instability of planetary waves was studied as a means to explain extratropical wavepacket formation. However, the geometry of the problem meant that zonal jet formation did not occur as a result. Manfroi \& Young \cite{manfroi} used the weakly nonlinear formulation of Sivashinsky \cite{sivvy} to study a more related problem involving the instability of a stationary planetary mode in the presence of bottom drag and viscosity. They found that asymmetric jets formed in their model, with an average separation that was dependent on the bottom drag. 
 
 The most relevant work to this section, however, is the previously cited Manin \& Nazarenko \cite{manin}, in which a Vlasov equation similar to (\ref{eq:WKstd}) was used to study the latitudinal instability of planetary waves in the limit of small $\beta$-effect. Here, we begin by summarising their methods and the result of their instability calculation. We then show how their calculation can be generalised using the results of Section \ref{sec:theory2}. It is found that inclusion of the mean flow correction terms results in a qualitatively different prediction for the fastest growing modes. The new instability predictions are then compared with results from the numerical model introduced in Section \ref{sec:compsim}. As in Section \ref{sec:compsim}, the vertical variation of all quantities is neglected here.

To reduce their mean flow equation into a tractable form, Manin \& Nazarenko assumed that the wavepacket density $\mathcal N$ could be treated as a $\delta$-function in wavenumber space dependent on a single dynamical variable $l(y,t)$
\begin{equation}
\mathcal N(\mathbf x, \mathbf k',t)=\mathcal N_0 \delta(k'-k_0) \delta[l' - l(y,t)].
\end{equation}
When this assumption is applied to (\ref{eq:waveaction_scalesep_gen}) and (\ref{eq:nonaccelbaro}) with $\kappa=0$, the result is
\begin{equation}\label{eq:modsimp1}
\frac{\partial \overline u}{\partial t} = \mathcal N_0 \frac{\partial v_y}{\partial y}.
\end{equation}
Equation (\ref{eq:modsimp1}) combined with the previous definition of `force' on a wavepacket
\begin{equation}\label{eq:modsimp2}
\frac{\partial l}{\partial t} = F_y =  -\frac{\partial \omega}{\partial y}
\end{equation}
then completely describes the evolution of the reduced system.

To find a dispersion relation for the modulational instability of a monochromatic wave, it is necessary to linearise (\ref{eq:modsimp1}) and (\ref{eq:modsimp2}) about a single wavenumber value $l=l_0+\tilde l$, $\overline u = \tilde u$, with $\tilde l, \tilde u \propto exp(-iK+i\Omega)$. In the work of Manin \& Nazarenko, the analysis was further simplified by earlier assumptions that a) the local planetary wave group velocity is only a function of $\beta$, not $\gamma$ and b) the Doppler term $k_0 \overline u$  (see equation (\ref{eq:Gdisp})) dominates all others in $\omega$.

In these circumstances, $v_y$ can be expanded in terms of $\tilde l$ only, and the dispersion relation for the modulational instability can be written as
\begin{equation}\label{eq:modcritmanin}
\Omega = \pm i K \sqrt{ \mathcal N_0 k_0 \left. \frac{\delta v_y}{\delta l}   \right|_{l=l_0} },
\end{equation}
where
\begin{equation}\label{eq:modcritmanin2}
 \left. \frac{\delta v_y}{\delta l}   \right|_{l=l_0} = \frac{2k_0\beta \left( k_0^2 - 3l_0^2 \right)}{\left(k_0^2+l_0^2 \right)^3}. \\
\end{equation}
Clearly, (\ref{eq:modcritmanin}) predicts the most unstable wave will always be the one of highest wavenumber $K$. It also predicts that a wave will only become unstable if 
\begin{equation}\label{eq:k0crit}
k_0^2 \slash l_0^2 > 3.
\end{equation}

 \begin{figure}
 	\begin{center}
 		{\includegraphics[width=3.5in]{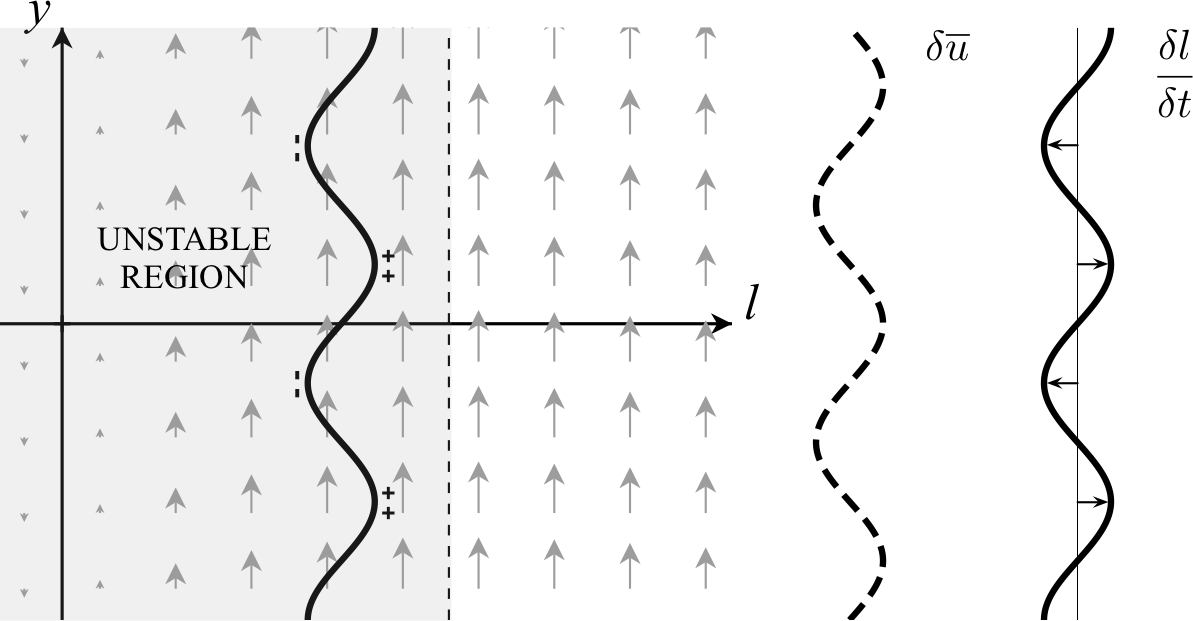}}
 	\end{center}
 	\caption{Phase-space interpretation of the instability criterion (\ref{eq:k0crit}). Grey arrows show local wavepacket velocity when no mean flow is present. The plus (minus) signs denote regions where wavepackets accelerate (decelerate) relative to the base group velocity $v_{0y}$. The mean flow response $\delta \overline u$ and resulting motion of wavepackets in the $l$-direction $\delta l \slash \delta t$ are also shown.}\label{fig:mod_cartoon}
\end{figure}

The instability criterion (\ref{eq:k0crit}) has an intuitive phase-space interpretation, which is visualised in Figure \ref{fig:mod_cartoon}. When a monochromatic planetary wave (equivalently, a thin phase-space strip of wave action) is perturbed in the $l$-direction, the gradient of base group velocity $v_{0y}$ (see Figure \ref{fig:beta_wp}) causes a convergence of wave action into certain regions and a divergence out of others. By the nonacceleration theorem, local wave action changes must cause an equal and opposite change in the local zonal velocity $\overline u$. In the unstable region of phase space (shaded gray in Figure \ref{fig:mod_cartoon}), the gradient of group velocity is such that this zonal velocity then causes the initial perturbation to grow via the shear effect discussed in Section \ref{sec:compsim} (see Figure \ref{fig:shear}). After the linear growth phase, the convergence of positive wave action into certain regions must cause the associated negative zonal jets to intensify. Thus jet asymmetry in the \emph{opposite sense} to that seen in Section \ref{sec:compsim} is expected to develop in this example.

\begin{figure}
 	\begin{center}
 		{\includegraphics[width=3.5in]{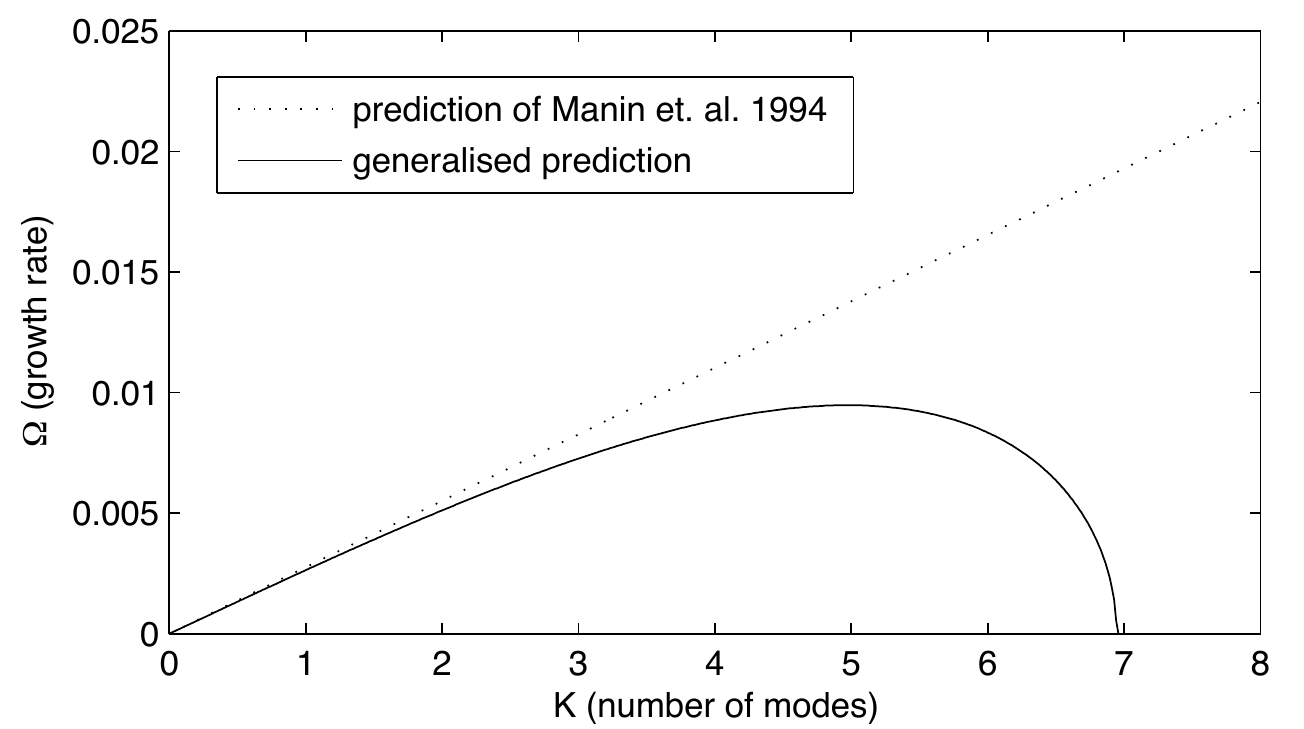}}
 	\end{center}
 	\caption{Growth rate as a function of mode number for the prediction of Manin \& Nazarenko (\ref{eq:modcritmanin}) and the generalised instability criterion (\ref{eq:modcritgen}).}\label{fig:mod_predict}
\end{figure}

If the two assumptions of Manin \& Nazarenko are not used, the instability analysis becomes more complicated. Multiple variable Taylor expansions of $\omega$ and $v_y$ yield
\begin{eqnarray}
i\Omega  \tilde u  &=& -iK \mathcal N_0\left( \left. \frac{\delta v_y}{\delta l}\right|_0 \tilde l +\left. \frac{\delta v_y}{\delta \overline u} \right|_0 \tilde u \right) \nonumber \\
i\Omega \tilde l    &=& +iK \left( \left. \frac{\delta \omega}{\delta \overline u}\right|_0  \tilde u  + \left. \frac{\delta \omega}{\delta l}\right|_0 \tilde l \right)
\end{eqnarray}
and hence
\begin{equation}\label{eq:modcritgen}
\Omega = \frac 12 K \left( V \pm  \sqrt{ V^2 - 4 \mathcal N_0\left(  \left.  \frac{\delta v_y}{\delta l}  \right|_0 \left. \frac{\delta \omega}{\delta \overline u}  \right|_0  -  v_{0y}    \left.  \frac{\delta v_y}{\delta \overline u}  \right|_0   \right) }\right).
\end{equation}
Here $V=v_{0y} - \mathcal N_0( \delta v_y \slash \delta \overline u)|_0 $ is the wavepacket velocity at $l=l_0$ with a correction due to mean flow effects. The explicit forms of the partial derivatives are
\begin{eqnarray}\label{eq:partialderivs}
\left. \frac{\delta \omega}{\delta \overline u}\right|_0  &=&   \frac{-k_0K^2}{k_0^2+l_0^2} + k_0 \qquad \left. \frac{\delta \omega}{\delta l} \right|_0 = v_{0y} = \frac{2k_0l_0 \beta}{(k_0^2+l_0^2)^2} \nonumber \\
\left. \frac{\delta v_y}{\delta \overline u}  \right|_0 &=&   \frac{2k_0l_0K^2}{(k_0^2+l_0^2)^2} \qquad \quad \left. \frac{\delta v_y}{\delta l} \right|_0 = \frac{2k_0 \beta (k_0^2 - 3l_0^2)}{(k_0^2+l_0^2)^3}
\end{eqnarray}
where $|_0$ implies evaluation at $l=l_0$ and $\overline u = 0$. The mean flow corrections in (\ref{eq:partialderivs}) are negligible for small $K$. At high $K$ values, however, they become increasingly important, with the result that imaginary part of $\Omega(K)$ has a definite maximum, after which it decreases to zero. In Figure \ref{fig:mod_predict}, (\ref{eq:modcritmanin}) and (\ref{eq:modcritgen}) are plotted as function of $K$ for an example where $k_0=50$, $l_0=6\pi$, $\beta=1$ and $\mathcal N_0=6\times10^{-4}$. As can be seen, their predictions diverge at high $K$ values, with the latter peaking at approximately $K=5$ modes. Equation (\ref{eq:modcritgen}) gives a maximum $\Omega(K)$ value at finite $K$ in a similar way for all cases where the criterion (\ref{eq:k0crit}) is satisfied.

To test the validity of this scale-separated analysis, we now compare its predictions with the results of numerical simulation. The setup is very similar to that used in \mbox{Section \ref{sec:compsim}}. One exception is the boundary conditions in the $y$-direction, which are now set to be periodic, for simplicity. Strictly speaking, the derivation of Section \ref{sec:theory2} is only valid for unbounded domains, but we expect that it should still work reasonably well as long as the wavelength of any energetic modes in the system are several times less than the latitudinal domain size, $\lambda_{max}<L_y$. In each simulation, the initial condition consists of a planetary wave of definite wavenumber $k_0$ and $l_0$, plus a small amount of random noise. The system is allowed to evolve until the Rayleigh-Kuo criterion $\beta-\partial_{yy}\overline u>0$ is violated, and then the wavenumber of the most energetic zonal mode is determined via a Fourier transform.

 \begin{figure}
 	\begin{center}
 		{\includegraphics[width=3.5in]{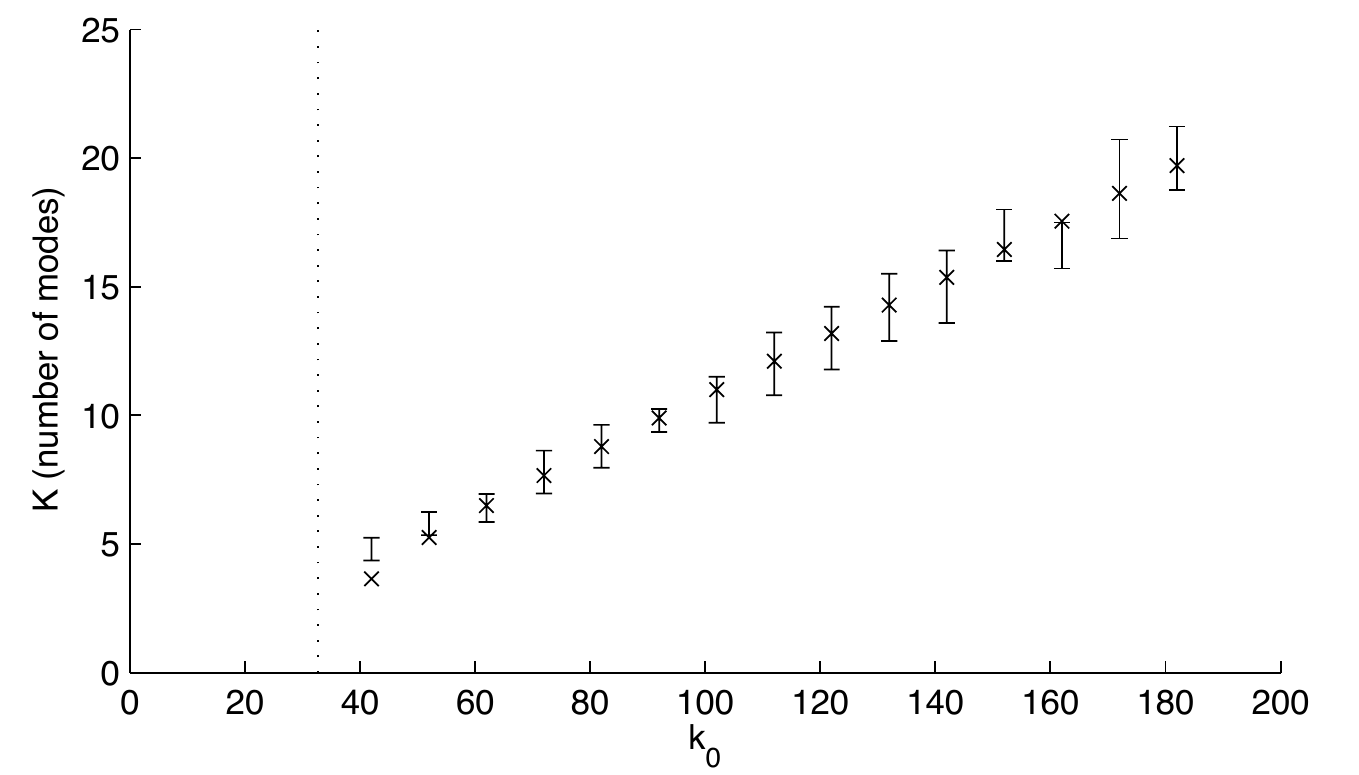}}
 	\end{center}
 	\caption{Theoretically predicted fastest growing mode (crosses) and most energetic zonal modes at onset of barotropic instability in simulation (error bars). Dotted line shows critical $k_0$ value as given by (\ref{eq:k0crit}).}\label{fig:mod_compare}
\end{figure}

In Figure \ref{fig:mod_compare}, the results of many such simulations, performed for varying values of $k_0$, are plotted alongside theoretical predictions. All other quantities were fixed at the same values as those for Figure \ref{fig:mod_predict}. As the largest mode in the simulation was found to vary unpredictably over a given range for each $k_0$ value, it was decided to perform five simulations at each point. It is the standard deviation about the average mode number that is plotted in Figure \ref{fig:mod_compare}. Resolution in the $y$-direction was set at $ny=128$; increasing this value did not significantly affect the results.

As can be seen, (\ref{eq:modcritgen}) accurately predicts the fastest growing mode at most values of $k_0$, with the variation nearly linear above the critical $k_0$ value. The reason for the slight divergence at low $k_0$ is not known; it is possible that the periodic boundary conditions or other non-local effects played a role in these cases. The dependence of the prediction on  $l_0$ was also tested; it was found that even for the extreme case $l_0=0$, the agreement between theory and simulation was fairly close.  Finally, the variation with $\mathcal N_0$ was tested, and it was found that both the predicted and simulated fastest growing modes remained almost constant with $\mathcal N_0$ over fairly large ranges.

\begin{figure}
 	\begin{center}
 		{\includegraphics[width=3.5in]{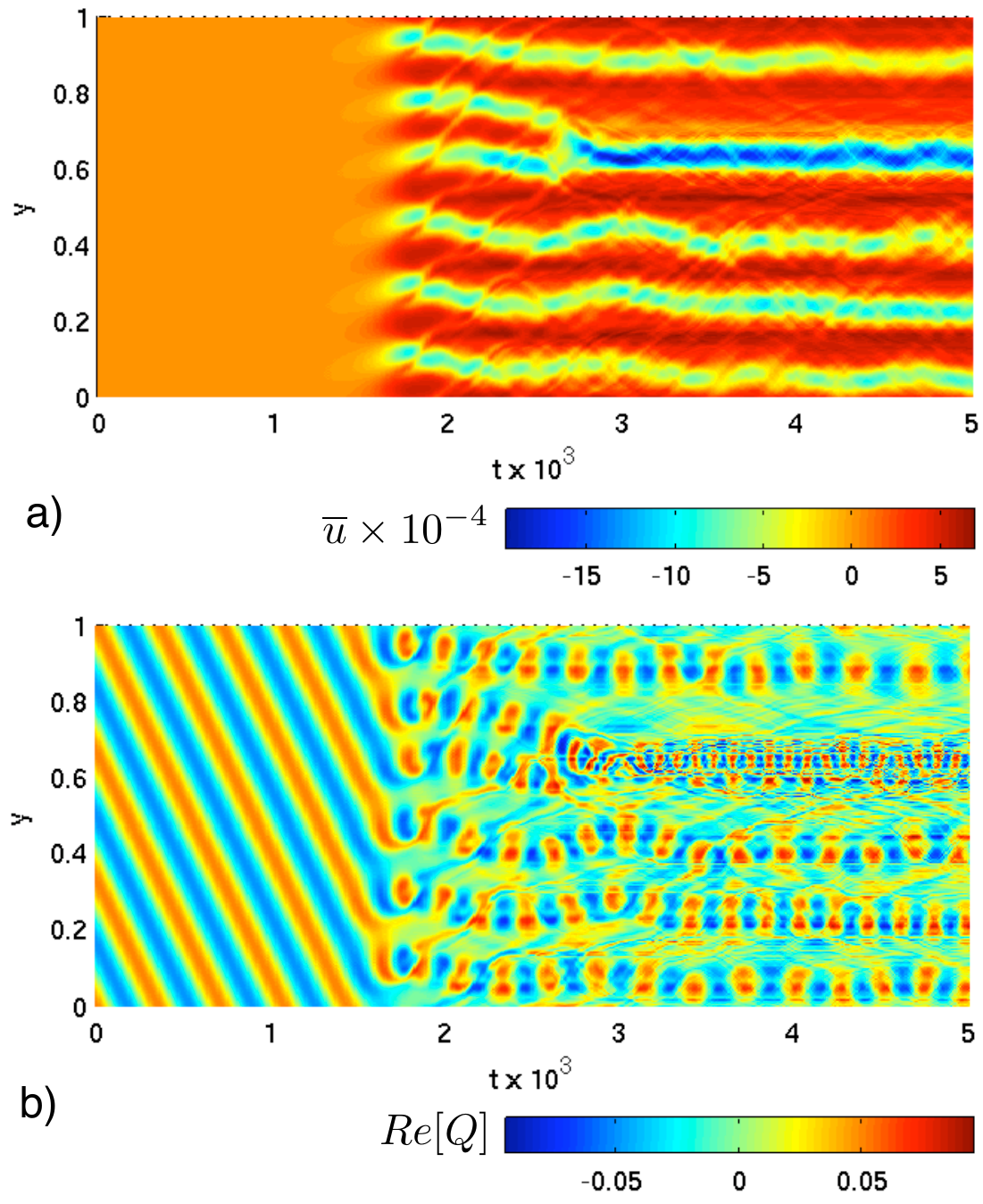}}
 	\end{center}
 	\caption{(Colour online) Space-time plot of a) zonal velocity $\overline u$ and b) disturbance vorticity component $Re[Q]$ for a modulational instability simulation with $k_0=50$, $l_0=6\pi$, $\mathcal N_0=1.25\times10^{-3}$ and $\beta=1$.}\label{fig:mod_pretty}
\end{figure}

In general, this analysis predicts a fastest growing zonal mode that is of \emph{larger} wavenumber than that of the base planetary wave, $l_0$. Because of this, and the linearity assumption introduced after (\ref{eq:modsimp2}), it cannot be expected to remain valid after the initial growth phase. As an example, Figure \ref{fig:mod_pretty} shows a space-time plot of $Re[Q]$ and $\overline u$ for a simulation that was allowed to continue for a long time after the onset of barotropic instability. As we are neglecting wave-wave interactions and restricting the wavefield to a single east-west wavenumber $k_0$, the results of this simulation cannot be assumed to be physically realistic. However, they were judged interesting enough to merit a brief discussion here.

The initial fastest growing zonal mode, which is clearly of different wavenumber from the base planetary wave, evolves rapidly after the barotropic instability criterion is broken (time $\approx 1750$). The expected east-west asymmetry quickly develops, and two of the growing jets can be seen to merge at time $\approx 2750$, with the resultant jet stronger than any of the others. The correlation between the zonal flow and the wavefield is extremely interesting; note that the jets appear to be trapping wavepackets, which propagate inside for the duration of the simulation in most cases. Calculation of the real-space wave action $n$ confirmed that the nonacceleration theorem was obeyed closely even to the end of the simulation.

Although idealised, this quasilinear study points the way towards more general investigations of jet formation via the modulational instability mechanism. In particular, it would be most interesting in future work to compare the predictions of (\ref{eq:modcritgen}) with a) a wave-mean flow simulation that allows multiple east-west wavenumbers and b) a fully nonlinear $\beta$-plane simulation.

\section{Discussion}\label{sec:conc}

We have studied the interaction between an arbitrary zonal flow and a broadband distribution of planetary waves. First, a new planetary-wave Vlasov equation was derived that included \emph{all} effects of the mean flow on the waves. A simple wave-mean flow numerical model was then used to investigate the interaction between planetary waves and zonal flow in a series of simple test cases. Jet formation and asymmetry were intuitively explained in terms of the motion of planetary wavepackets in phase space. 

Next, the quasilinear instability of an initially monochromatic planetary wave was investigated. A generalisation of the analysis of Manin \& Nazarenko \cite{manin} was used to predict a finite fastest growing zonal wavenumber. This was then compared with the results of a number of numerical simulations for varying values of east-west wavenumber $k_0$. The prediction was found to give good agreement with numerical results for a wide range of parameter values. Jet asymmetry was again observed in the simulation (although in the opposite sense to that in the wavepacket example) and explained via a phase-space argument that made use of the results of Section \ref{sec:compsim}.

The results presented here have demonstrated the effectiveness of the Wigner-based approach to wave-mean flow analysis. While previous studies have noted the possibility of $\beta$-plane modulational instability and estimated a timescale for the growth rate when it occurs, this appears to be the first work in which a fastest growing zonal mode for an arbitrary planetary wave is predicted and the results tested against a more general numerical simulation.

It is quite possible that planetary-wave modulational instability is an important mechanism for multiple jet formation in real planetary atmospheres and oceans. However, many of the assumptions used in this paper do not apply in more realistic scenarios. To investigate the importance of the mechanism further, therefore, it would be interesting to generalise the analysis presented here in several ways.

One of the main assumptions made in Vlasov equation derivations is that scale separation exists between the mean flow and waves. Some progress has been made in generalising beyond this assumption. For example, in Powell et~al. \cite{powell}, the scattering of waves off random topography in the absence of a mean flow was studied using an expansion of a Wigner transport equation to 1\textsuperscript{st} order. They successfully used their method to derive transport equations for planetary waves propagating in a two-layer model with random topography included. Unfortunately, the presence of a finite-amplitude zonal flow in the system appears to make progress in this direction more difficult. 

Given the correspondence between the scale-separated predictions and more general numerical simulation in this paper, however, this generalisation may not be of primary importance. Perhaps more limiting is the restriction to wave-mean flow interactions only, which was used throughout the analysis. Wave-wave and turbulent interactions can play a very important role in the overall development of fluid flows on the $\beta$-plane, and any eventual general theory should aim to take them into account. 

This generalisation may not be easy either, as standard techniques for the derivation of wave-kinetic equations appear inapplicable in situations where the zonal flow is of finite amplitude \cite{manin}. It is possible that an extension of the operator-based derivation of Section \ref{sec:theory2} could be used to tackle this problem. As a first step, however, it would be interesting and relatively simple to compare the theoretical and numerical results presented here with a fully nonlinear numerical simulation.

\begin{acknowledgments}
The author would like to thank his supervisor, Prof.  P.~L.~Read, and also Prof. D.~G.~Andrews, for helpful suggestions and advice on a draft version of this paper. The Wigner distribution plots were created using a modified version of a \emph{Matlab} script, the original of which is freely available online at \emph{http://page.mi.fu-berlin.de/burkhard/WavePacket/}. This work was funded by a Natural Environments Research Council studentship.
\end{acknowledgments}

\bibliographystyle{plain}

\end{document}